\documentclass[12pt,epsf,epsfig]{article}
\usepackage{amsmath,latexsym}

\begin{document}

\begin{center}
{\huge{Superstrings with the Galileon Measure.}} \\
\end{center}

\begin{center}
T.O. Vulfs \textsuperscript{1,2,3}, E.I. Guendelman \textsuperscript{1,3,4} \\
\end{center}

\begin{center}
\textsuperscript{1} Department of Physics, Ben-Gurion University of the Negev, Beer-Sheva, Israel \\
\end{center}

\begin{center}
\textsuperscript{2} Fachbereich Physik der Johann Wolfgang Goethe Universit\"{a}t, Campus Riedberg, Frankfurt am Main, Germany \\
\end{center}

\begin{center}
\textsuperscript{3} Frankfurt Institute for Advanced Studies, Giersch Science Center, Campus Riedberg, Frankfurt am Main, Germany \\
\end{center}

\begin{center}
\textsuperscript{4} Bahamas Advanced Study Institute and Conferences, 4A Ocean Heights, Hill View Circle, Stella Maris, Long Island, The Bahamas \\
\end{center}

E-mail: vulfs@post.bgu.ac.il, guendel@bgu.ac.il

\abstract
The modified measure theories recommend themselves as a good possibility to go beyond the standard formulation to solve yet unsolved problems. The Galileon measure that is constructed in the way to be invariant under the Galileon shift symmetry is considered in the context of superstring theory. The translation invariance of the vacuum holds up to a Galileon transformations. The supersymmetric action is presented with all terms, including the tension, being derived from the equations of motion.

\section{Introduction}
The action of a physical system is a Lagrangian density, $L$, integrated over the D-dimensional spacetime. In General Relativity the invariant volume element is usually constructed out of the tensor densities, namely, the determinant of the metric, $g$, and the volume element, $d^D x$: $\sqrt{-g}d^D x$. However, its modifications carry us beyond the standard theory. A long-standing problems become solvable. This paper is devoted to Galileon measure theory (GMT) \cite{d} and its big brother - Two measure theory (TMT) \cite{e}. \\
The former is originated from the Galileon modification of gravity \cite{a,b,c}. It is a scalar-tensor theory with a nonminimal coupling of the special scalar Galileon to curvature, with the second order equations of motion and nonetheless, without ghosts. This theory is invariant under a Galileon shift symmetry. This very symmetry is a base for this measure. \\
The later provides a thought out answers for the cosmological constant problem \cite{f}, the emergent universe scenario \cite{g}, the fifth force problem \cite{h}, the Dark Energy/Dark Matter scenarios \cite{n}. \\
Due to the evident success in gravity, we apply here the modified measures to String Theory \cite{o,a1,a2,a3}. Beside the general incompleteness, the theory of all matter and interactions contains the dimensionfull parameter - the tension of the string. This problem is contemplated in TMT and GMT. \\
When including the fermions, the supersymmetry is required. In the Green-Schwarz formulation \cite{k} the Wess-Zumino term is invariant only up to a total divergence. However, in Siegel reformulation \cite{l} this term becomes manifestly supersymmetric. But the price to pay is that now it consists of some vector fields that in principle are not determined by the equations of motion. It was considered in TMT \cite{m}. \\
In this paper we address these problems of naturally undetermined quantities to GMT. \\
In Section 2 we review the similarities and differences between TMT and GMT. The derivation of the tension as an integration constant is presented in Section 3. Section 4 is devoted to the standard formulation of the superstring action. In Section 5 we show the main result - the superstring action - with all the meaningfull and exactly supersymmetric terms. The discussion and conclusions are provided in Section 6. \\

\section{Modified Measure Theories}

The physical system is characterized by the action, $S$. It is not unique and it could be written in several different forms that depend on one's purposes. Strings being one-dimensional objects sweep out a two-dimensional surface in spacetime. By the direct geometric analogue with the point particle, the proper area of this surface is indeed an action. The Nambu - Goto action is

\begin{equation}
S_{Nambu-Goto} = -T \int d\tau d\sigma \sqrt{-h},
\end{equation}

where $h$ is the determinant of the induced metric and $\tau$ and $\sigma$ are the parameters on the worldsheet. \\

However, for the further quantization the metric under the square root is not welcome. Therefore, the new formulation directed on the elimination of $\sqrt{-h}$ appeared. It was accompanied by the introduction of the intrinsic metric, $\gamma^{ab}$, that is a dynamical variable in $S$. The sigma-model action is

\begin{equation} \label{eq:05}
S_{sigma-model} = -T\int d\tau d\sigma \sqrt{-\gamma} \gamma^{ab}\partial_a X^{\mu} \partial_b X^{\nu} g_{\mu\nu},
\end{equation}

where $g_{\mu\nu}$ is the metric on the embedding spacetime. \\

This action is also called the Polyakov action because Polyakov was the first to use it for the path integral quantization of the string. But it was Deser and Zumino and independently Brink, Di Vecchia and Howe who actually proposed it. \\

The sigma-model action has its own field equations that show explicitly how $h$ and $\gamma$ are related. \\

Both formulations contain $T$, the string tension. It brings a scale in the otherwise scale invariant theory. \\

This is the reason to reformulate the action again. Our alterations affect the integration measure, $\sqrt{-\gamma}$. We are limited only to measures that are densities under diffeomorphic transformations, $\Phi$.

\subsection{TMT}

The measure $\Phi(\varphi)$ is constructed out of two (the number of dimensions) scalar fields, $\varphi_i, \varphi_j$ and it does not depend on any metric:

\begin{equation}
\Phi(\varphi) = \epsilon_{ij} \epsilon^{ab} \partial_a\varphi^i \partial_b\varphi^j \qquad \rightarrow \qquad S = \int d\tau d\sigma \Phi(\varphi) L,
\end{equation}

where $L$ is arbitrary. \\

The variation with respect to $\varphi_i$ is

\begin{equation}
\epsilon^{ab} \partial_b \varphi_j \partial_a L= 0.
\end{equation}

Then, if $\Phi(\varphi)$ is not degenerate, it leads to the condition:

\begin{equation} \label{eq:01}
L = const.
\end{equation}

\subsection{GMT}

The measure $\Phi(\chi)$ is constructed out of one scalar field $\chi$ and do depend on the metric:

\begin{equation}
\Phi(\chi) = \partial_h (\sqrt{-\gamma} \gamma^{hd} \partial_d \chi) \qquad \rightarrow \qquad S = \int d\tau d\sigma \Phi(\chi) L.
\end{equation}

The scalar field $\chi$ is a Galileon since $\Phi(\chi)$ is invariant under a Galileon shift symmetry in the conformally flat metric gauge:

\begin{equation}
\partial_{a}\chi \rightarrow \partial_{a}\chi + b_{a},
\end{equation}

\begin{equation}
\chi \rightarrow \chi +b_{a}\sigma^{a},
\end{equation}

where $b_a$ is a constant vector and $\sigma^a = (\tau, \sigma)$. \\

This symmetry is exact and available only in two dimensions, for only in two dimensions there exist a conformally flat frame for $\gamma^{hd}$. \\

We should point out that a shift of the scalar by a linear function of coordinates resembles a "gauge symmetry" (although the gauge function is restricted to be a linear function of the coordinates, not a general function). These kind of "scalar gauge fields" were considered in \cite{b1,b2}. \\

The variation with respect to $\gamma^{ab}$, that is the density of the energy-momentum tensor, is

\begin{equation}
T^{ab} = \frac{-2}{\sqrt{-\gamma}} \frac{\delta (L \Phi(\chi))}{\delta \gamma_{ab}}.
\end{equation}

Then we obtain after some calculations:

\begin{equation}
T^{ab} = -\frac{2}{\sqrt{-\gamma}} \frac{\partial L}{\partial \gamma_{ab}}\Phi -\partial^a \chi \partial^b L - \partial^b \chi \partial^a L + \gamma^{ab} \partial_e \chi \partial^e L.
\end{equation}

The trace equation is

\begin{equation}
\gamma_{ab}T^{ab} = \frac{2}{\sqrt{-\gamma}} L \Phi(\chi) = 0,
\end{equation}

where we have used that $L$ is homogeneous of degree 1 in $\gamma^{ab}$. \\

It leads to the constraint (if $\Phi \ne 0$):

\begin{equation} \label{eq:02}
L = 0.
\end{equation}

Relations (\ref{eq:01}) and (\ref{eq:02}) determine the connection point between these modified measure theories. \\

\section{Effects of Modified Measure Theories}

The dynamics of the physical system is characterized by equations of motion. Regardless of $S$ formulations, they are always the same. It is a criterion of validity for any new theory. However, their amount which depends on the dynamical variables may differ. \\

The dynamical variables of the sigma-model action (\ref{eq:05}) are $\gamma^{ab}$, $X^{\mu}$. \\

The corresponding equations of motion are

\begin{equation} \label{eq:06}
T_{ab} = (\partial_a X^{\mu} \partial_b X^{\nu} - \frac12 \gamma_{ab}\gamma^{cd}\partial_cX^{\mu}\partial_dX^{\nu}) g_{\mu\nu}=0,
\end{equation}

\begin{equation} \label{eq:07}
\frac{1}{\sqrt{-\gamma}}\partial_a(\sqrt{-\gamma} \gamma^{ab}\partial_b X^{\mu}) + \gamma^{ab} \partial_a X^{\nu} \partial_b X^{\lambda}\Gamma^{\mu}_{\nu\lambda}=0,
\end{equation}

where $\Gamma^{\mu}_{\nu\lambda}$ is the affine connection for the external metric. \\

Now we consider theories with modified measures. We have defined $\Phi(\varphi)$ and $\Phi(\chi)$ and the restriction on $L$ in the previous section. In this section we construct $L$ itself. \\

The simplest proposition is

\begin{equation}
L_{simple} = \gamma^{ab}\partial_a X^{\mu} \partial_b X^{\nu} g_{\mu\nu},
\end{equation}

It fails in TMT because then $\Phi(\varphi)$ or the induced metric must vanish. The same conclusion follows in GMT because of the constraint (\ref{eq:02}). \\

However, any term that is a total derivative (when multiplied by a measure) could be added to $L$ without any consequences for the equations of motion. The measure is modified, the preference for the complemental term is changed. \\

The contribution is

\begin{equation} \label{eq:08}
L_{additional} = \frac{\epsilon^{ab}}{\sqrt{-\gamma}} F_{ab},
\end{equation}

where $F_{ab} = \partial_a A_b - \partial_b A_a$ is the field-strength of an auxiliary Abelian gauge field, $A_a$. \\

The relation (\ref{eq:08}) is obtained from general considerations, satisfies the conditions above and induces the consistent results in what follows. \\

Then the action is

\begin{equation} \label{eq:03}
S = -\int d \tau d \sigma \Phi (L_{simple} + L_{additional}),
\end{equation}

where $\Phi$ could be either $\Phi(\chi)$ or $\Phi(\varphi)$.

\begin{equation} \label{eq:04}
L_{simple} + L_{additional} = const \qquad in \qquad TMT,
\end{equation}

\begin{equation}
L_{simple} + L_{additional} = 0 \qquad in \qquad GMT.
\end{equation}

The TMT action is

\begin{equation}
S_{TMT} = -\int d^2\sigma (\gamma^{cd}\partial_c X^{\mu} \partial_d X^{\nu} g_{\mu\nu} - \frac{\epsilon^{cd}F_{cd}}{\sqrt{-\gamma}})\epsilon^{ab} \epsilon^{ij} \partial_a\varphi_i \partial_b\varphi_j.
\end{equation}

The GMT action is

\begin{equation}
S_{GST} = -\int d^2\sigma (\gamma^{cd}\partial_c X^{\mu} \partial_d X^{\nu} g_{\mu\nu} - \frac{\epsilon^{cd}F_{cd}}{\sqrt{-\gamma}}) \partial_h (\gamma^{hd} \sqrt{-\gamma} \partial_d \chi).
\end{equation}

Variations of (\ref{eq:03}) with respect to $\gamma^{ab}$ and $X^{\mu}$ and the conformal symmetry considerations which determine the constant in (\ref{eq:04}) lead to the same equations of motion, (\ref{eq:06}), (\ref{eq:07}). \\

But now there is an extra dynamical variable, $A_a$. It is the variation with respect to the gauge field that is responsible for the appearance of $T$ as an integration constant:

\begin{equation}
\epsilon^{ab} \partial_b (\frac{\Phi}{\sqrt{-\gamma}}) = 0.
\end{equation}

Then again (if $\Phi \ne 0$):

\begin{equation} \label{eq:09}
\frac{\Phi}{\sqrt{-\gamma}} = Const.
\end{equation}

The correspondence principle tells us that this const is $T$. Then

\begin{equation} \label{123123123}
\frac{\Phi}{\sqrt{-\gamma}} = T.
\end{equation}


\section{The Vacuum Solution}

The local conformal invariance of the equations of motion was discussed in \cite{d}. So an intrinsic metric can be considered as a flat one: $\gamma^{ab} = \eta^{ab}$. It leads to $\Box = \partial_a(\eta^{ab}\partial_b)$. Then from (\ref{123123123}) the following equation is obtained

\begin{equation}
\Box\chi = T.
\end{equation}

The general solution is

\begin{equation} \label{321321321}
\chi = \frac{T}{8} \eta_{ab} \sigma^a \sigma^b + c_b x^b + c + \phi_h,
\end{equation}

where $c_b$ and $c$ are some constant vector and scalar correspondingly and $\phi_h$ satisfies the sourceless equation $\Box \phi_h = 0$. The checking is

\begin{equation}
\partial_a(\eta^{ab} \partial_b (\frac{T}{8} \eta_{cd} \sigma^c \sigma^d)) = \partial_a(\eta^{ab} \frac{T}{8} \eta_{cd}(\delta^c_b \sigma^d + \delta^d_b \sigma^c)) = T.
\end{equation}

The relation(\ref{321321321}) is taken to be the vacuum. It is not translation invariant under $\sigma^a \rightarrow \sigma^a + b^a$, where $b^a$ is some constant vector. However, it possesses a translation invariance up to a Galileon transformation. Namely, (\ref{321321321}) transforms as $\phi \rightarrow \phi + d_bx^a + d$, where $d_b$ and $d$ are some constant vector and scalar correspondingly. \\

So the discussion of translation invariance of the solution (\ref{321321321}) is similar to the one on the translation invariance of the magnetic monopole under rotations  \cite{qaz}. The rotational invariance holds up to a gauge transformation. \\

The 't Hooft - Polyakov monopole obviously is spherically symmetric. However, the effect of a rotation is non-trivial and must be compensated by a corresponding rotation in isospin space. Here by analogy, the translation invariance must be made into an exact symmetry by combining it with a Galileon symmetry. \\

\section{Standard Measure Superstrings}

The previous sections concern only the bosonic strings. The inclusion of fermions requires the supersymmetry. Among two standard approaches, the Ramond - Neveu - Schwarz and the Green - Schwarz ones, we choose the later by the reason of the manifest spacetime supersymmetry provided. In addition to $X^{\mu}(\tau, \sigma)$ there are Grassmann-valued coordinates, $\theta^{\alpha}(\tau, \sigma)$, where $\alpha$ is a two-dimensional spinor index. \\

The supersymmetry transformations are

\begin{equation}
\delta \theta^{\alpha} = \epsilon^{\alpha},
\end{equation}

\begin{equation}
\delta X^{\mu} = -i(\epsilon \Gamma^{\mu} \theta),
\end{equation}

where $\Gamma^{\mu} \equiv \Gamma^{\mu}_{\alpha \beta}$ are the Dirac matrices. \\

By the analogue with a point particle, the supersymmetric modification is

\begin{equation}
\Pi^{\mu}_{\alpha} = \partial_{\alpha} X^{\mu} - i(\theta \Gamma^{\mu} \partial_{\alpha} \theta).
\end{equation}

Then the action is

\begin{equation}
S = -T \int d\tau d\sigma \sqrt{-\gamma} L_{simpleSUSY},
\end{equation}

where

\begin{equation}
L_{simpleSUSY} = \frac12 \sqrt{-g} g^{\alpha \beta} \Pi^{\mu}_{\alpha} \Pi_{\beta \mu}.
\end{equation}

Even the invariance under global transformations does not guarantee that this $S$ is a complete solution. \\

A local fermionic symmetry (the kappa symmetry) must be preserved. Then the so-called Wess-Zumino term is

\begin{equation}
L_{additionalSUSY} = i\epsilon^{\alpha \beta} \partial_{\alpha} X^{\mu}(\bar{\theta^1} \Gamma_{\mu} \partial_{\beta} \theta^1 + \bar{\theta^2} \Gamma_{\mu} \partial_{\beta} \theta^2) -\epsilon^{\alpha \beta} \bar{\theta^1} \Gamma^{\mu} \partial_{\alpha} \theta \bar{\theta^2} \Gamma_{\mu} \partial_{\beta} \theta^2.
\end{equation}

The Green-Schwarz superstring action is

\begin{equation}
S_{Green-Schwarz} = -T \int d\tau d\sigma \sqrt{-\gamma}(L_{simpleSUSY} + L_{additionalSUSY}).
\end{equation}

Both $L_{simpleSUSY}$ and $L_{additionalSUSY}$ are invariant under local reparametrizations but opposite to $L_{simpleSUSY}$, $L_{additionalSUSY}$ is invariant under global transformations only up to total derivatives. \\

The next formulation bases on the idea of constructing the action via the supersymmetric currents. The Siegel action is

\begin{equation} \label{eq:10}
S_{Siegel} = -T \int d\tau d\sigma \sqrt{-\gamma} (\frac12 \gamma^{ab} J^{\mu}_a J^{\nu}_b \eta_{\mu\nu} + i\frac{\epsilon^{ab}}{\sqrt{-\gamma}}J^{\alpha}_a J_{\alpha b}),
\end{equation}

where

\begin{equation}
J_a^{\alpha} = \partial_a \theta^{\alpha},
\end{equation}

\begin{equation}
J_a^{\mu} = \partial_a X^{\mu} - i(\partial_a \theta \Gamma^{\mu} \theta),
\end{equation}

\begin{equation}
J_{\alpha a} = \partial_a \phi_{\alpha} - 2i(\partial_a X^{\mu}) \Gamma_{\mu \alpha \beta} \theta^{\beta} - \frac23 (\partial_a \theta^{\beta})\Gamma^{\mu}_{\beta \delta} \theta^{\delta} \Gamma_{\mu \alpha \epsilon} \theta^{\epsilon}.
\end{equation}

The global symmetry is exact now, not just up to a total divergence. It is $\phi_{\alpha}$ that allows the Wess-Zumino term to be expressed in the manifestly supersymmetric way. \\

Then additional transformation is

\begin{equation}
\delta \phi_{\alpha} = 2i\epsilon^{\beta} \Gamma_{\mu \alpha \beta} X^{\mu} + \frac23 (\epsilon^{\beta} \Gamma^{\mu}_{\beta \epsilon} \theta^{\epsilon}) \Gamma_{\mu \alpha \kappa} \theta^{\kappa}.
\end{equation}

However, the $\phi_{\alpha}$ fields are not dynamical. \\

The modified measure theories have just proved themselves extremely useful in giving meaning to the sudden constants.

\section{Galileon Measure Superstrings}

We have already addressed to the gauge field, $A_a$, to modify the string action. By doing it again, we define

\begin{equation}
-i \epsilon^{ab} \partial_a \theta^{\alpha} \partial_b \phi_{\alpha} = \epsilon^{ab} \partial_a A_b,
\end{equation}

\begin{equation} \label{eq:11}
A_b = -i \theta^{\alpha} \partial_b \phi_{\alpha}.
\end{equation}

Then the superstring with the Galileon measure is

\begin{equation} \label{eq:12}
S_{GMTSUSY} = -\int d\tau d\sigma (\frac12 \gamma^{ab} J^{\mu}_a J^{\nu}_b \eta_{\mu\nu} + i\frac{\epsilon^{ab}}{\sqrt{-\gamma}}J^{\alpha}_a J_{\alpha b})\partial_h (\gamma^{hd} \sqrt{-\gamma} \partial_d \chi).
\end{equation}

The metric $\gamma^{ab}$ does not participate in supersymmetric transformations. \\

The constraint is already inside the action. Namely,

\begin{equation}
\frac12 \gamma^{ab} J^{\mu}_a J^{\nu}_b \eta_{\mu\nu} + i\frac{\epsilon^{ab}}{\sqrt{-\gamma}}J^{\alpha}_a J_{\alpha b} = 0.
\end{equation}

Equations of motion are the same as obtained previously for (\ref{eq:10}), the consistency is proved. \\

However, what matters, we have the additional equation of motion that determines $\phi_{\alpha}$. This same result was derived in TMT but as in the string case the steps were different. \\

The variation with respect to $\phi_{\alpha}$ is

\begin{equation}
\epsilon^{ab} \partial_a \theta^{\alpha} \partial_b (\frac{\Phi(\chi)}{\sqrt{-\gamma}}) = 0.
\end{equation}

Therefore, in the nondegenerate case $(\partial_a \theta^{\alpha} \ne 0)$, we have

\begin{equation}
\frac{\Phi(\chi)}{\sqrt{-\gamma}} = const.
\end{equation}

The result is very similar to the one obtained for the bosonic string (\ref{eq:09}). Then again the constant is a tension of the string. \\

Another formulation of supersymmetric strings is also possible. It was obtained in the framework of TMT in \cite{i}. This action resembles the Siegel action if the gauge field is taken to be (\ref{eq:11}). \\

When applying to the Galileon measure, we obtain

\begin{equation} \label{eq:13}
S_{GMTSUSY} = \int d\tau d \sigma (-\frac12 \gamma^{ab} \Pi^{\mu}_a \Pi_{b \mu} + \frac{\epsilon^{ab}}{\sqrt{-\gamma}}(\Pi^{\mu}_a (\theta \Gamma_{\mu} \partial_b \theta) + \frac12 F_{ab})) \partial_h (\gamma^{hd} \sqrt{-\gamma} \partial_d \chi).
\end{equation}

Equations of motion are identical, the legitimacy is confirmed.

\section{Discussion and Conclusion}

We have derived the supersymmetric string action with a Galileon measure, (\ref{eq:12}), (\ref{eq:13}). At every step we were guided by symmetry principles. The integration measure is a density under diffeomorphisms on the worldsheet. It is constructed in such a way as to possess the Galileon shift symmetry. Moreover, when considering an action, the measure does not break the conformal transformation symmetry. \\
Since $\Phi(\chi)$ is a total derivative, $L \rightarrow L + const$ is a symmetry. The ability to add (\ref{eq:08}) is fateful for the whole theory. \\
The very idea of including fermions to the bosonic string totally reposes on the supersymmetry. \\
Further developments could concern other formulations of supersymmetric strings. Yet, the enumerated symmetries should be preserved. \\

\textbf{Acknowledgments}
TV acknowledges support by the Ministry of Aliyah and Integration (IL). EG is supported by the Foundational Questions Institute.

\end{document}